\documentclass{ws-mpla}
\usepackage[super,compress]{cite}
\usepackage{graphicx}
\usepackage{epsfig}% Include figure files
\begin{document}
\markboth{Monojit Ghosh, Srubabati Goswami and Sushant K. Raut}{Implications of $\delta_{CP}=-90^\circ$ towards determining hierarchy and octant at T2K and T2K-II}

\def\be{\begin{equation}}
\def\ee{\end{equation}}
\def\bea{\begin{eqnarray}}
\def\eea{\end{eqnarray}}\def\nn{\nonumber}
\def\gsim{\ \rlap{\raise 2pt\hbox{$>$}}{\lower 2pt \hbox{$\sim$}}\ }
\def\lsim{\ \rlap{\raise 2pt\hbox{$<$}}{\lower 2pt \hbox{$\sim$}}\ }
\def\dslash{\kern-4pt \not{\hbox{\kern-2pt $\partial$}}}
\def\pslash{\not{\hbox{\kern-2pt p}}}
\def\p{{\bf p}}
\def\q{{\bf q}}
\def\gev{{\rm GeV }}
\def\l{{\rm L}} 
\def\db{{$\delta b$ }}
\def\dc{{$\delta c$ }}

\def\evsq{{${\rm eV^2}$ }}
\def\L{{\rm L }}
\def\E{{\rm E }}

\def\pmutau{{{P_{\mu \tau}} }}
\def\pmue{{{P_{\mu e}} }}
\def\pmumu{{{P_{\mu \mu}} }}
\def\pemu{{{P_{e \mu}} }}
\def\pee{{{P_{ee}} }}

\def\numutonutau{{${\rm {\nu_\mu \to \nu_\tau}}$ }}
\def\numutonue{{${\rm {\nu_\mu \to \nu_e}}$ }}
\def\numutonumu{{${\rm {\nu_\mu \to \nu_\mu}}$ }}

\def\nuetonutau{{${\rm {\nu_e \to \nu_\tau}}$ }}

\def\pnumutonutau{{${\rm P ({\nu_\mu \to \nu_\tau})}$ }}
\def\pnumutonue{{${\rm P ({\nu_\mu \to \nu_e})}$ }}
\def\pnumutonumu{{${\rm P ({\nu_\mu \to \nu_\mu})}$ }}

\newcommand{\dcp}{\delta_{CP}}
\newcommand{\nova}{NO$\nu$A}
\newcommand{\dms}{\Delta m^2_\odot}
\newcommand{\dma}{\Delta m^2_{\rm atm}}
\newcommand{\dmsq}{\Delta m^2}
\newcommand{\ahat}{\hat{A}}

\catchline{}{}{}{}{}
%
%%%%%%%%%%%%%%%%%%%%%%%%%%%%%%%%%%%%%%%%%%%%%%%%%%%%%%%%%%%%%%%%%%%%

\title{Implications of $\dcp=-90^\circ$ towards determining hierarchy and octant at T2K and T2K-II}

\author{Monojit Ghosh}

\address{Department of Physics, Tokyo Metropolitan University, Hachioji, Tokyo 192-0397, Japan\\
monojit@tmu.ac.jp}

\author{Srubabati Goswami}

\address{Physical Research Laboratory, Navrangpura,
Ahmedabad 380 009, India\\
sruba@prl.res.in}

\author{Sushant K. Raut}

\address{Department of Theoretical Physics, School of Engineering Sciences,
KTH Royal Institute of Technology -- AlbaNova University Center,
Roslagstullsbacken 21, 106 91 Stockholm, Sweden\\
raut@kth.se}

\maketitle

\begin{history}
\received{Day Month Year}
\revised{Day Month Year}
\end{history}

\begin{abstract}
The T2K experiment has provided 
the first hint for the best-fit value  for the 
leptonic CP phase $\delta_{CP} \sim -90^\circ$  from neutrino data. 
This is now corroborated by the \nova~ neutrino runs. 
We study the implications for neutrino mass hierarchy and octant of 
$\theta_{23}$ in the context of this data assuming
that the true value of $\dcp$ in nature is $-90^\circ$.
Based on simple arguments on degeneracies in the probabilities we show that 
a clear signal of 
$\dcp=-90^\circ$
coming from T2K neutrino (antineutrino) data is only possible 
if the true hierarchy is normal and the true
octant is higher (lower). 
Thus if the T2K neutrino and antineutrino data are fitted separately and both give the true value of $\dcp=-90^\circ$, 
this will imply that nature has chosen the true hierarchy to be normal
and $\theta_{23} \approx 45^\circ$. 
However we find that the combined fit of neutrino and antineutrino data will 
still point to
true hierarchy as normal but the octant of $\theta_{23}$ will remain
undetermined. We do our analysis for both, the current projected exposure 
($7.8 \times 10^{21}$ pot) and planned extended exposure ($20 \times 10^{21}$
pot). 
We also  present the CP discovery potential 
of T2K emphasizing on 
the role of antineutrinos.
We find that one of the main contribution of the antineutrino data
is to remove the degenerate solutions with the wrong octant. 
Thus the antineutrino run plays a more significant role  for those   
hierarchy-octant combinations for which this degeneracy 
is present. If this degeneracy is absent, then 
only neutrino run gives a better result for fixed $\theta_{13}$. 
However if we marginalize over $\theta_{13}$ then, sensitivity 
corresponding to mixed run can be better than pure neutrino run. 

\end{abstract}

\ccode{PACS numbers:14.60.Pq}

%\tableofcontents

\section{Introduction}

Neutrino physics is currently poised at an interesting juncture.
Among the parameters of the neutrino mass matrix, 
oscillation experiments have 
measured the mass squared differences $\Delta_{21}$, $|\Delta_{31}|$ 
($\Delta_{ij} = m_i^2 - m_j^2$) and the mixing angles 
($\theta_{12},\theta_{23},\theta_{13}$) 
with considerable precision.
The remaining 
unknown oscillation parameters  are
(i) the mass hierarchy: 
normal hierarchy (NH, $m_3 > m_2 > m_1$) or inverted hierarchy
(IH, $m_3 < m_2 \approx m_1$),  
(ii) octant of $\theta_{23}$: $\theta_{23} < 45^\circ$(lower octant, 
LO) or $\theta_{23} > 45^\circ$ (higher octant, HO) and 
(iii) the CP-violating phase $\dcp$.
The global analysis of current oscillation data gives no  statistically 
significant indication of the mass hierarchy.
There is also no clear indication of the octant 
of $\theta_{23}$. For inverted hierarchy higher octant is preferred 
while for normal hierarchy the lower octant is preferred
\cite{global,fogli,valle}. 
However these indications are still fragile.   
Recently, an indication for $\delta_{CP} \sim -90^\circ$ has been 
obtained by a combination of T2K  and reactor data~\cite{t2krecent}. This hint
comes from  
T2K running in the neutrino mode with  
8\% of the expected total flux of T2K 
($7.8 \times 10^{21}$ protons on target (pot))~\cite{t2krecent}. 
The first results  from $\nu_\mu-\nu_e$ search from 
\nova\ have also shown a similar signal~\cite{Adamson:2016tbq}.
The best-fit value for $\dcp$ is obtained to be close to   $-\pi/2$, but at
$3\sigma$ C.L. the whole range of $[0,2\pi]$ is still admitted 
\cite{global,fogli,valle}.

%In this paper we consider the possibility of determination of the 
%above unknowns in future runs of T2K. 
The relevant channel to determine the current unknowns in the neutrino oscillation sector is 
the $\nu_\mu \to \nu_e$ oscillation probability $P_{\mu e }$ as 
it is sensitive to all 
the three unmeasured parameters described above. 
However the lack of knowledge of $\dcp$ 
can give rise to three additional spurious solutions 
corresponding to 
wrong hierarchy-right octant, right hierarchy-wrong octant and 
wrong hierarchy-wrong octant in addition to the correct solution.
Thus there is a 4-fold degeneracy which is a subset of the 
8-fold degeneracy discussed in the literature ~\cite{degen8fold}. 
Most of these degenerate solutions can occur for different values of 
$\delta_{CP}$ other than the true value making its determination 
difficult~\cite{Ghosh:2015ena}. 
 In this work, we study what the true value $\dcp=-90^\circ$
will imply for  
neutrino mass hierarchy and octant of $\theta_{23}$ in the context of
the T2K experiment taking its (i) full expected exposure ($7.8 \times 10^{21}$ 
pot per year   
and (ii) also the proposed enhanced exposure ($20\times 10^{21}$ pot per year 
\cite{Abe:2016tii}).

%This is motivated by the fact that the recent analysis of T2K data give a hint about the true value of $\dcp$ to be around $-90^\circ$.
% In that case, we argue that the data will also indicate the true hierarchy as 
% NH and true octant as HO. This is because, 
% for the other combinations of true hierarchy and octant in nature, 
% degenerate solutions occur 
% for wrong values of $\dcp$,
% preventing a distinct signature for 
% $\dcp = -90^\circ$. 
% This provides  an indirect indication 
% of the hierarchy and octant from T2K. 
% 
% Thus, for the first time we show that the 
% T2K experiment by itself can give a hint of 
% all three unknowns by running only in neutrino mode.
 In this context we also  elucidate on the role played by antineutrinos in
 improving CP sensitivity, and identify the cases for which antineutrino 
 runs can be helpful. 
An early hint of these unknown parameters from the 
T2K first phase of runs will be useful from the point of view 
of planning neutrino facilities in future. In addition, knowledge of 
these parameters provides an important test for neutrino mass models 
and will therefore significantly influence our search for models of 
new physics beyond the Standard Model.

% identify the cases for which only neutrino 
%runs will suffice for discovering CP violation in T2K. 
% discuss the effect of combined neutrino-antineutrino run of 
%T2K for various combinations of hierarchy and octant.  

\section{Degeneracies in $P_{\mu e}$}

The T2K experiment uses the neutrino beam from J-PARC and has a baseline
of 295 km. 
The probability relevant for the measurement of CP violation is 
\cite{akhmedov,lindner,Freund}, 
\bea
P_{\mu e }&=& 4 s_{13}^2 s_{23}^2 \frac{ \sin^2{(\ahat-1)\Delta} }{(\ahat-1)^2} \\ \nonumber
&+& 2 \alpha s_{13} \sin{2\theta_{12}} \sin{2\theta_{23}}\cos{(\Delta + \dcp)} \frac{\sin{\ahat\Delta}}{\ahat} \frac{ \sin{(\ahat-1)\Delta} }{\ahat-1}  + {\cal{O}}(\alpha^2) ~. 
\label{P-emu}
\eea
Here 
$s_{ij} (c_{ij}) \equiv \sin{\theta_{ij}}(\cos{\theta_{ij}})$,  
$\Delta = \Delta_{31}L/4E$ where $L$ is the distance traveled and $E$ is the energy of the neutrino. 
$\ahat = 2EV/\Delta_{31}$, where $V = \sqrt{2} G_F n_e$ is
Wolfenstein's matter potential in terms of the electron density $n_e$. 
%The leading term in the probability has octant sensitivity 
%(correlated with $\theta_{13}$) and 
%sub-leading term has hierarchy and $\dcp$ sensitivity.
The lack of definite information about hierarchy, octant and 
$\dcp$ gives rise to degenerate solutions. For our purpose we consider 
the following degeneracies:  
(i) Hierarchy-$\dcp$ degeneracy: $P_{\mu e}(\dcp, \Delta) = P_{\mu e}(\dcp^\prime, -\Delta^\prime)$ 
i.e. the probability for NH can be mimicked by IH and  a different    
$\dcp$ value giving rise to 
wrong hierarchy-wrong $\dcp$ solutions~\cite{degen8fold,degen,burget,minakata} 
and(ii) Octant-$\dcp$ degeneracy : $P(\theta_{23}^{LO},\delta_{CP})
= P(\theta_{23}^{HO},\delta'_{CP})$, i.e. the probability in
the right octant can be the same as that for wrong octant and a 
different $\dcp$ giving  
wrong octant-wrong $\dcp$ solutions~\cite{lisi,generalizeddegen,ghosh,coloma}.
%Both these degeneracies affect the hierarchy, octant and  CP sensitivity
%of T2K. 
Hierarchy determination is facilitated if nature has 
chosen favourable combinations of hierarchy and $\dcp$ 
--
\{$\dcp \in [-180^\circ,0^\circ]$, NH\} and 
\{$\dcp \in [0^\circ,180^\circ]$, IH\}), which hold for both neutrinos and 
antineutrinos~\cite{novat2k}. 
The situation is different for octant determination.
For true LO, the degenerate solutions arise for 
(\{$\dcp \in [-180^\circ,0^\circ]$ and for true HO they occur
for \{$\dcp \in [0^\circ,180^\circ]$\} in the neutrino mode; 
for antineutrino mode the behaviour is 
opposite~\cite{uma,machado,prakash}. 
This feature in the oscillation probability can be understood 
from the following simple arguments. 
For neutrinos the values of $P_{\mu e }$ are
higher for NH and lower for IH, and it is opposite for antineutrinos. But 
there is also a flip in the relative sign of $\dcp$ between neutrinos 
and antineutrinos.
That causes the hierarchy-$\dcp$ degeneracy to appear in the same region for 
both neutrinos and antineutrinos. On the other hand 
the value of $P_{\mu e }$ is lower for LO and higher for HO  
for neutrinos as well as antineutrinos. Therefore the octant-$\dcp$
degeneracy behaves differently with neutrinos and antineutrinos.
This implies that combination of neutrino and antineutrino
channel is helpful for removal of octant-$\dcp$ degeneracy 
but it does not help in removal of hierarchy-$\dcp$ degeneracy.

\begin{figure*}[ht]
  \includegraphics[width =\textwidth]{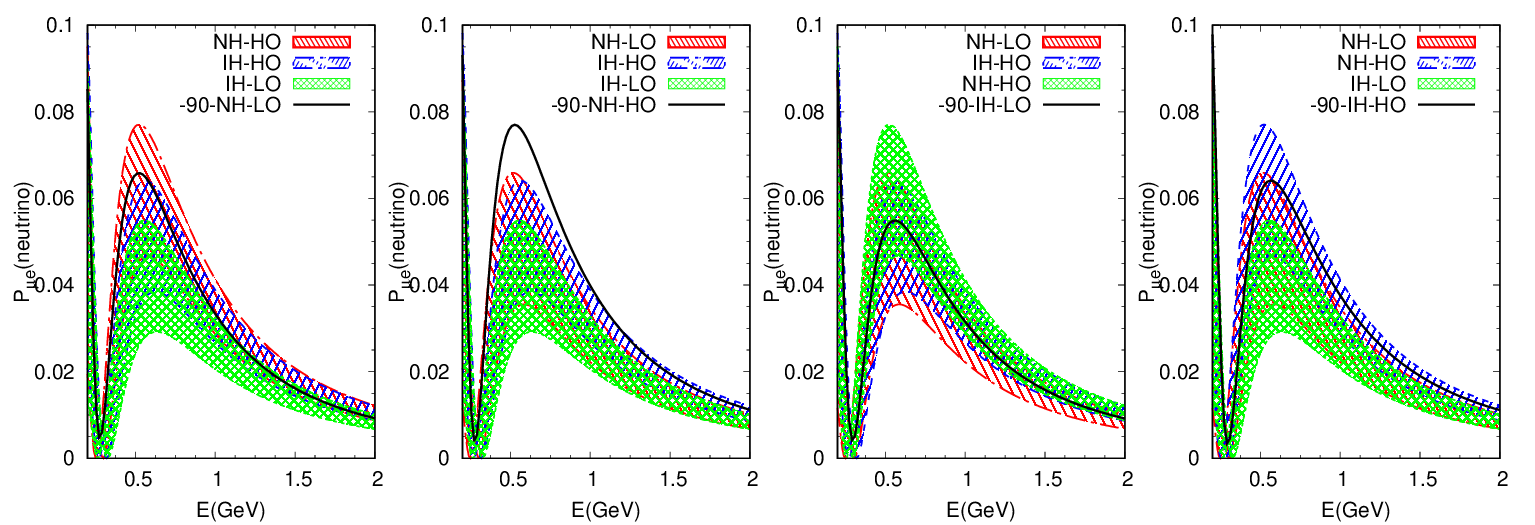}
  \includegraphics[width =\textwidth]{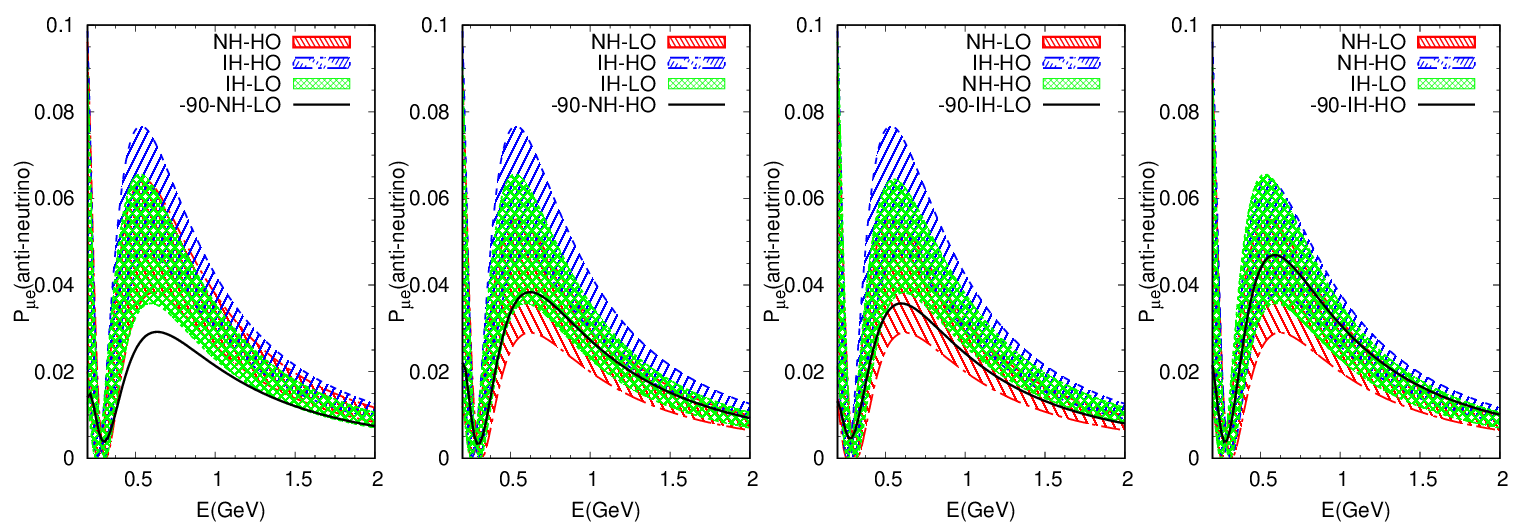}
  \includegraphics[width =\textwidth]{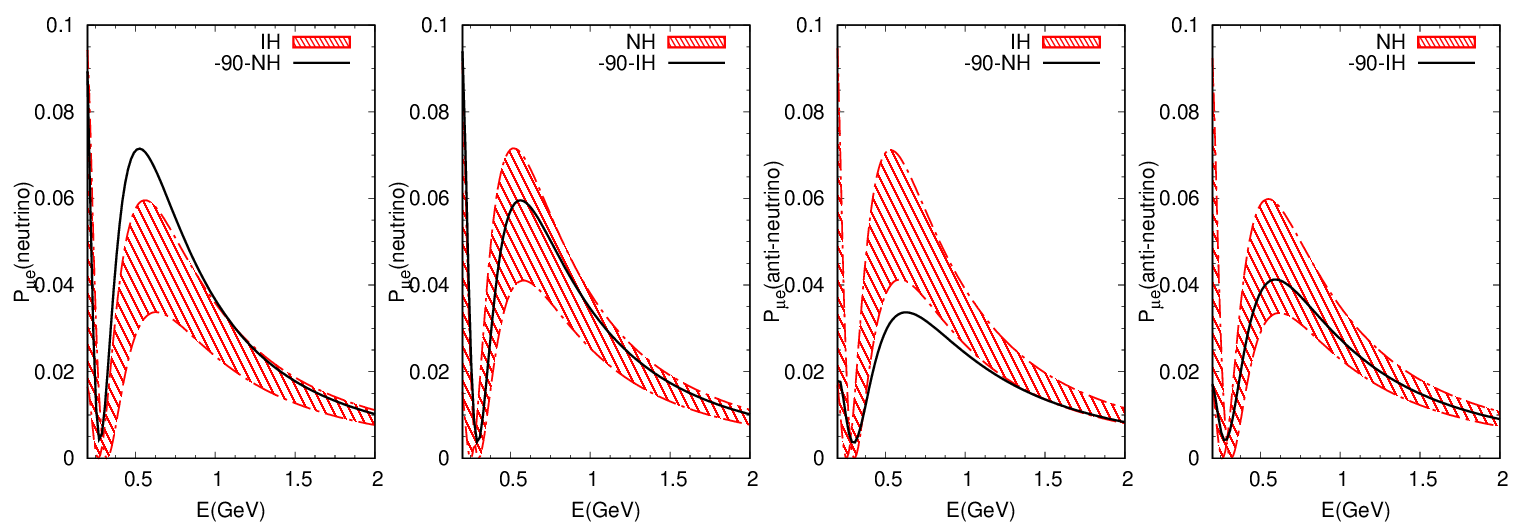}
  \caption{Probability figures showing degeneracy for $\dcp=-90^\circ$ at T2K baseline. 
  Upper (middle) row is for neutrinos (antineutrinos). The bands are due to the variation of $\theta_{23}$ and $\dcp$.
  The lower row is for $\theta_{23}=45^\circ$ and the bands are due to the variation of $\dcp$.
  }
  \label{prob}
  \end{figure*}
  
  \begin{figure*}[ht!]
  \includegraphics[width =\textwidth]{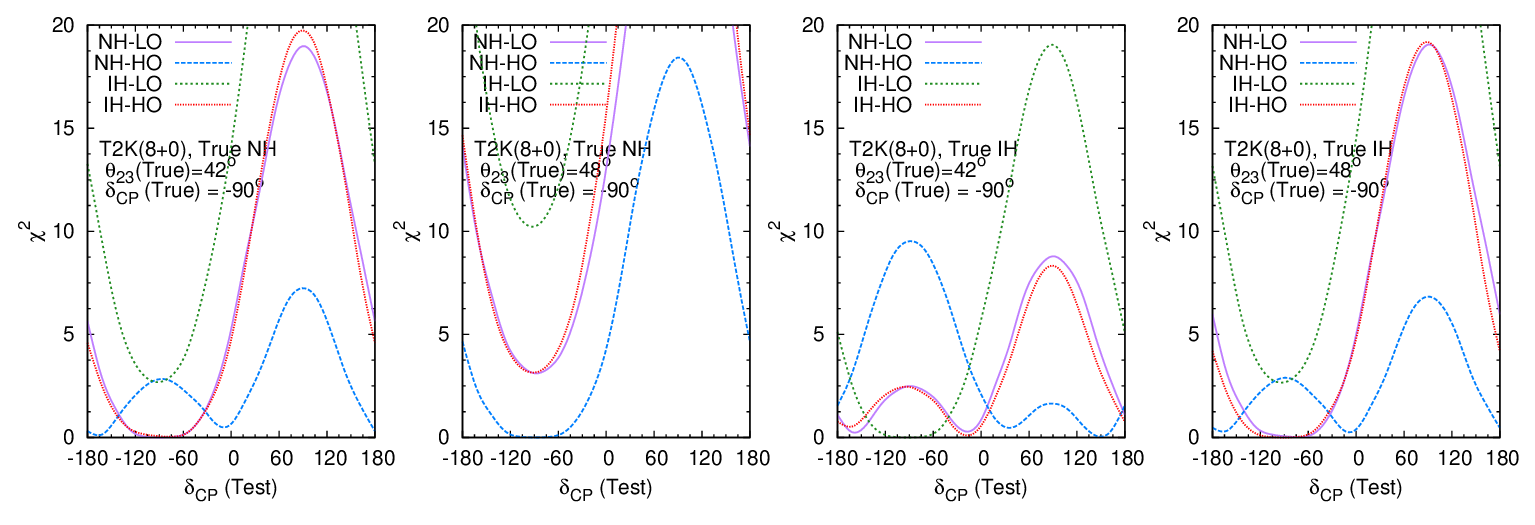}
  \includegraphics[width =\textwidth]{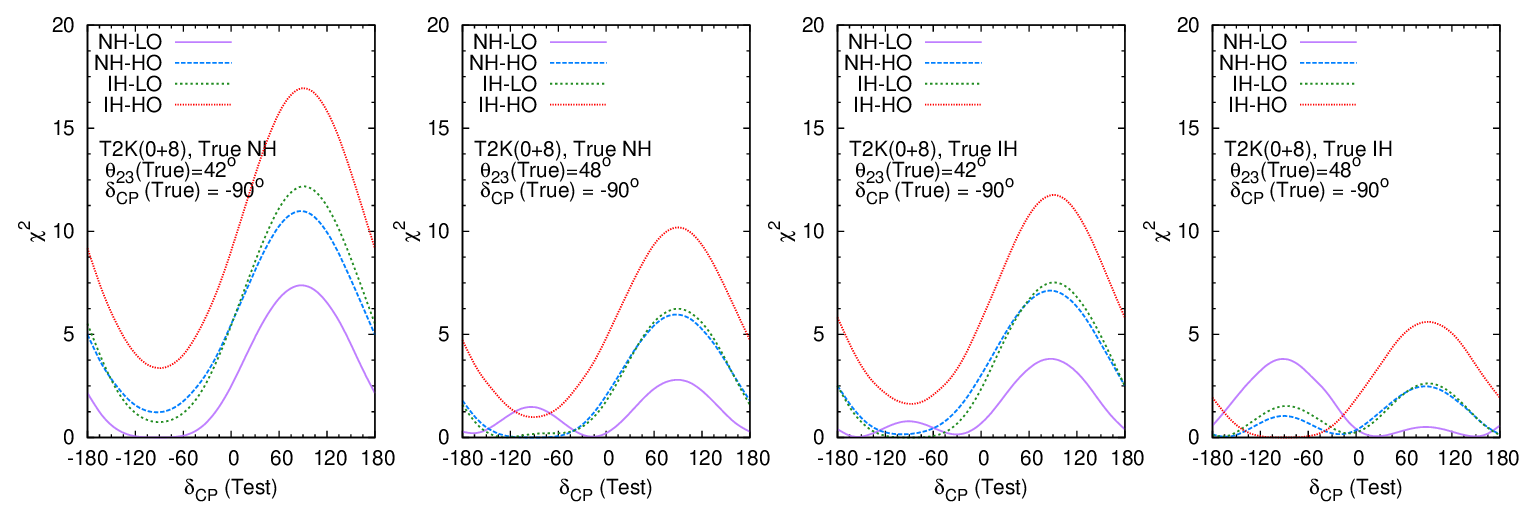}
  \includegraphics[width =\textwidth]{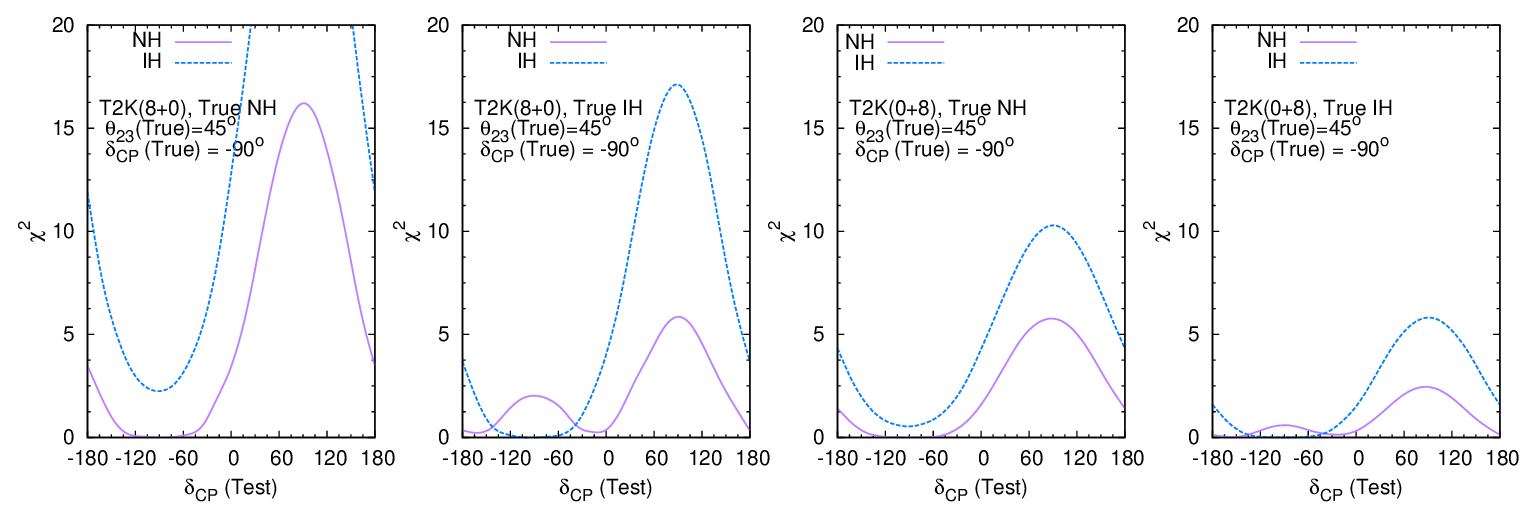}
  \includegraphics[width =\textwidth]{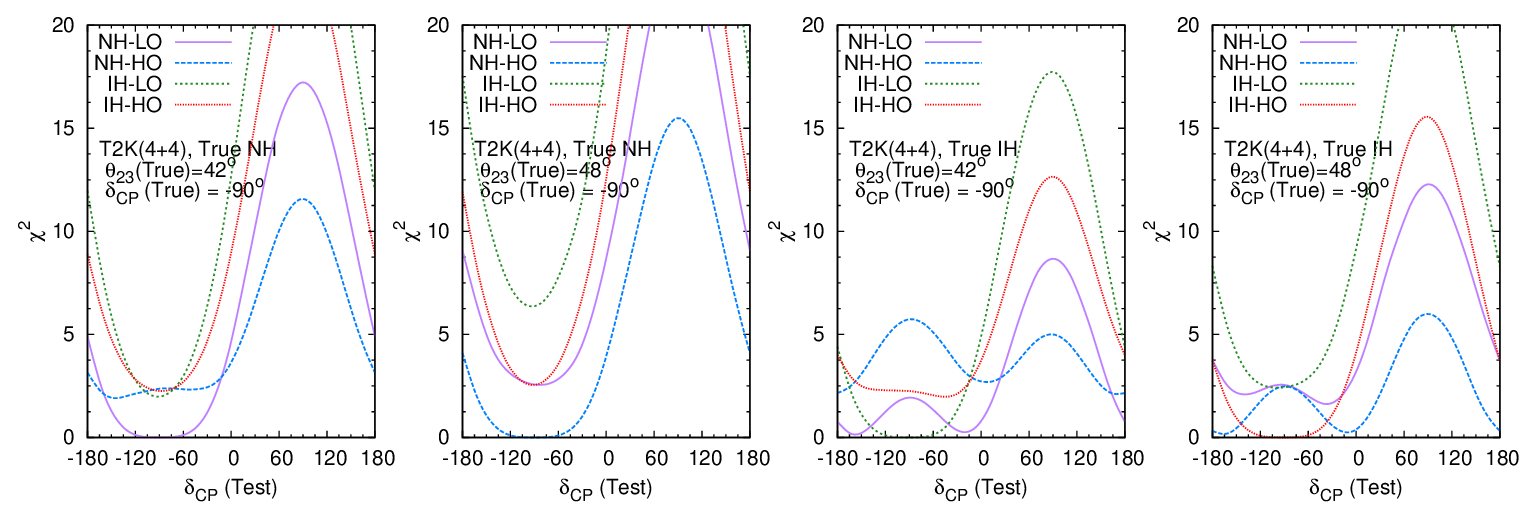}
  \caption{Hint for hierarchy and octant assuming true value of $\dcp=-90^\circ$ for an exposure of $8 \times 10^{21}$ pot. The first (second) row is for pure neutrinos (antineutrinos).
  In the third row $\theta_{23}$ is $45^\circ$ and in the fourth row, an equal neutrino+antineutrino run of T2K is assumed.
  }
  \label{cpnurun}
  \end{figure*}

To understand the degeneracy at $\dcp=-90^\circ$, in Fig. \ref{prob} we plot the appearance channel probability $P_{\mu e}$ vs energy for the T2K baseline.
In the upper panels we show the probabilities for the neutrinos 
for $\dcp = -90^\circ$ (solid black line) for four different combinations 
of true hierarchy and octant. 
For instance in the first panel the black solid line corresponds to 
NH-LO. Also plotted are the probability bands (obtained by varying over the octant of $\theta_{23}$ and
CP phase)  for the other 
three combinations.  The figure shows that for $\dcp= -90^\circ$ and NH-LO
there is complete degeneracy in the probability with the other cases. 
Thus a clear signature of $\dcp = -90^\circ$ like the one 
seen by T2K would be difficult to obtain. 
Similar comparisons are made for the other three true hierarchy octant 
combination. A comparison across panels shows that only for NH-HO the 
probability for $\dcp=-90^\circ$ is free of any degeneracy and can 
result in a clear signal at T2K. 
In all other cases there will be degenerate solutions. 
Thus from probability level discussion on degeneracies
it is clear that a signal for $\dcp = -90^\circ$ 
in the neutrino channel will imply the hierarchy-octant combination to be
NH-HO. 
In the middle panels of Fig. \ref{prob} we present similar figures for the 
antineutrino channel. 
From this figure it is clear that excepting true NH-LO in all other cases there 
will be degenerate solutions. 
Thus just from probability level discussion it is apparent that a signal for $\dcp=-90^\circ$ 
in neutrino channel implies the octant to be higher whereas for antineutrinos
it has to be lower.  Thus if both neutrino and antineutrino channel imply 
$\dcp=-90^\circ$ then there is an ambiguity about the octant. 
Can this imply that the mixing angle $\theta_{23}$ is maximal ?. 
To check this in the lower panels of Fig. \ref{prob}, we plot the probabilities for $\dcp = -90^\circ$ assuming the 
true hierarchy to be NH and IH for both neutrinos and antineutrinos. 
The red shaded band shows the probability obtained by varying $\dcp$ for the 
opposite hierarchy. In this case we see that for NH there is 
no degeneracy and the true curve is separated from the red shaded curve 
for both neutrinos and antineutrinos. 
Note that the above discussion is at the
probability level. In the actual scenario, sensitivity of an experiment depends upon
the  event rates and associated statistical and systematic errors. In the next section
we present a $\chi^2$ level representation of the above results including these.

\section{Results}

We simulate the T2K experiment using 
the GLoBES package~\cite{globes,globes1} along with its 
auxiliary files~\cite{globes_aux,globes2,globes3,globes4,globes5,globes6,globes7}. 
We consider T2K running  
a total of $7.8 (20) \times 10^{21}$ pot. 
We will use the notation $(a+b)$ to denote T2K runtime throughout the paper. The number $a$ and $b$ denotes exposure of T2K in units of $10^{21}$ pot in neutrino and antineutrino mode
respectively.
Event rates for both appearance and disappearance channels have been simulated for various combinations of hierarchy 
(NH or IH) and octant (LO or HO). Here LO corresponds to $\theta_{23}=42^\circ$ 
and HO corresponds to $\theta_{23}=48^\circ$, while NH/IH correspond to 
$\Delta_{31} = \pm 2.4\times10^{-3} \textrm{eV}^2$. The true 
value of $\dcp$ in nature is taken to be $-90^\circ$ unless specified otherwise. 
Note that the issue of the best-fit value and octant 
of $\theta_{23}$ is an intriguing one. The T2K data gives the 
best-fit $\theta_{23}$ as $45^\circ$~\cite{Abe:2015awa}  while \nova~ data gives two 
statistically indistinguishable  best-fit 
points one in the LO and the other in HO for both hierarchies~\cite{Adamson:2016xxw}.   
The allowed range of  $\theta_{23}$ from the latest global analysis is  
$38^\circ - 53^\circ$~\cite{nufit} for any hierarchy.  
Considering all these aspects, in our analysis we take three representative 
values of $\theta_{23}$ -- $42^\circ$ (LO), $48^\circ$ (HO) and  $45^\circ$
(maximal mixing). It is to be noted that the octant sensitvity will be more for 
$\theta_{23}$ further  away from the maximal value.  
The true values and marginalization ranges of the oscillation parameters used in 
our analysis are listed in Table~\ref{tab:oscvals}.

\begin{table}[ph]
\tbl{Table summarizing the values and ranges of oscillation parameters used for the sensitivity study.}
{ \begin{tabular}{|c|c|c|}
  \hline
  Parameter & True value & Marginalization range \\
  \hline
  $\theta_{12}$ & $33^\circ$ & fixed \\
  $\sin^22\theta_{13}$ & 0.1 & 0.085 - 0.115 \\
  $\theta_{23}$ & \begin{tabular}{c} $42^\circ$(LO) \\ $45^\circ$(maximal) \\ $48^\circ$(HO) \end{tabular} & $38^\circ-52^\circ$ \\
  $\Delta_{21}$ & $7.5 \times 10^{-5} \textrm{eV}^2$ & fixed \\
  $|\Delta_{31}|$ & $2.4 \times 10^{-3} \textrm{eV}^2$ & $(2.19-2.61) \times 10^{-5} \textrm{eV}^2$ \\
  $\dcp$ & as specified & $-180^\circ - +180^\circ$ \\
  \hline
 \end{tabular}
%\caption{Table summarizing the values and ranges of oscillation parameters used for the sensitivity study.}
\label{tab:oscvals} }
\end{table}

% We call this 8 T2K-year which corresponds to $10^{21}$ pot/year. 
% Note that this is not 8 year in real time because of the initial 
% low power runs of T2K. 

\subsection{Hint for hierarchy and octant}
\subsubsection{Results for $7.8 \times 10^{21}$ pot}

In Fig.~\ref{cpnurun}, we have plotted the sensitivity of T2K to measure $\dcp$. 
The first row corresponds to 
pure neutrino run of T2K. 
In the first panel of first row, NH-LO is taken as the true combination of hierarchy and octant. The various 
curves show fits to $\dcp$ for all of the four test combinations, of which one is 
correct and the other three wrong. Expectedly, the correct combination (NH-LO) gives 
a best-fit (minimum $\chi^2$) at the true value of $\dcp=-90^\circ$. With NH-HO as 
the test combination, we get a best-fit around $180^\circ$ while with IH-HO as the 
test combination, we have a best-fit close to $-90^\circ$. The combination IH-LO is 
seen to be excluded with minimum $\chi^2 \approx 2.5$. 
Therefore, on marginalizing over the
hierarchy and octant to find the overall 
best-fit value of $\dcp$, one would see allowed values of $\dcp$ around $-90^\circ$, 
and $180^\circ$. In other words, there would not be a strong indication 
for any single value of $\dcp$ from the data. 
Similar conclusions can be drawn if the 
true combinations in nature are IH-LO and IH-HO (third and fourth panel).
% For  true IH-HO (first row fourth panel) the additional minima can come at
% $\dcp = 0 and \pm 180^o$ for NH-HO.  
% Thus if we assume that the  CP conserving case is 
% not allowed then only 
% a clear hint at $-90^o$ can be obtained.} 
The true combination NH-HO is the only one, 
for which one can see an unambiguous 
signal at $-90^\circ$, as the second panel shows. This is easy to 
explain, since neutrino probabilities for NH are higher than for IH, 
and those for 
HO are higher than those for LO. Therefore, it is not possible for any other 
combination of parameters to match the high event rates of NH-HO and 
create a degeneracy when $\dcp = -90^\circ$. 
Thus, a hint for $\dcp=-90^\circ$  would also signify normal mass 
hierarchy and higher
octant of $\theta_{23}$ by elimination of the other options.

\begin{figure*}
  \epsfig{file=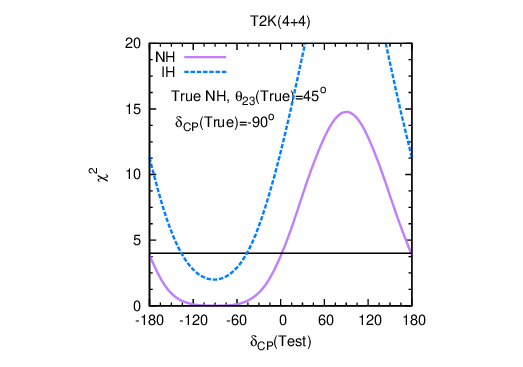,  width=0.45\textwidth, bbllx=79, bblly=50, bburx=270, bbury=255,clip=} 
  \epsfig{file=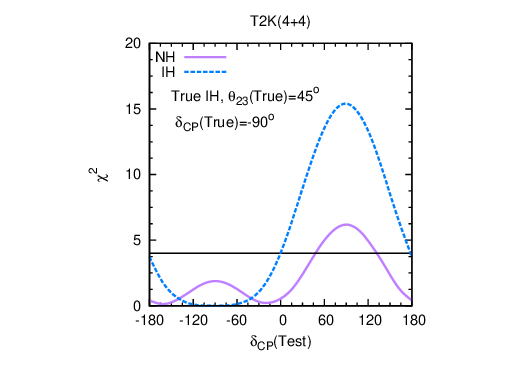,  width=0.45\textwidth, bbllx=79, bblly=50, bburx=270, bbury=255,clip=} 
  \caption{$\dcp$ sensitivity of T2K for equal neutrino+antineutrino run for maximal value of $\theta_{23}$ for an exposure of $7.8 \times 10^{21}$ pot.}
  \label{cpantinurun}
  \end{figure*}

Now let us come to the second row of Fig. \ref{cpnurun} which corresponds to the pure antineutrino run of T2K.
Following the same arguments made above we can see that except NH-LO (first panel), it is not possible
measure a single value of $\dcp$ for the other true combinations hierarchy and octant. This is because those combinations have 
other best-fit points apart from $\dcp=-90^\circ$ 
\footnote{It is worth emphasizing here that NH-LO is favoured not by fitting 
the hierarchy and octant as shown in the first panel alone (since the other cases are 
disfavoured at very low statistical confidence), but by recognizing that in all other 
true cases (i.e. comparing all four panels) one would not get the unambiguous 
signal for $\dcp=-90^\circ$ that the experiments have already seen.}. 
This is also easy to understand from the probability arguments. For antineutrinos
$-90^\circ$-NH-LO correspond to lowest value in the probability spectrum and 
thus free from any degeneracy. This can be seen from Fig.~\ref{prob}. 

From the above discussion we understand that if nature chooses $\dcp=-90^\circ$ as the true value of $\dcp$ then measuring it via fitting the
neutrino data will imply the true combination of hierarchy and octant as NH-HO and if it is measured via fitting the antineutrino data then it will be NH-LO.
Thus there is an inconsistency regarding the octant of $\theta_{23}$ except if nature has chosen $\theta_{23}$ to be maximal. To see this, in the third row of
Fig. \ref{cpnurun} we have plotted the CP sensitivity of T2K assuming $\theta_{23}=45^\circ$. From the figure we can see that for both pure neutrino and pure antineutrino 
cases only in the case of 
true NH it is possible to have an unambiguous determination of $\dcp=-90$ (first and third panel) whereas true IH gives multiple best-fit values of $\dcp$
apart from $\dcp=-90^\circ$.

\begin{figure*}[ht!]
  \includegraphics[width =\textwidth]{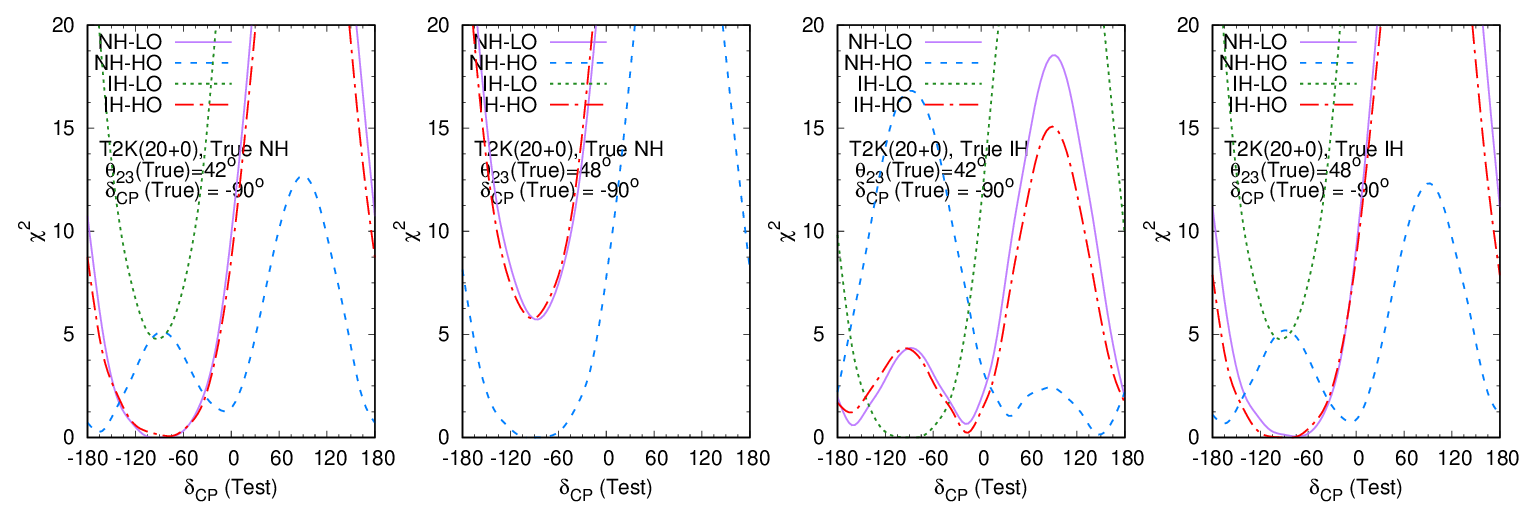}
  \includegraphics[width =\textwidth]{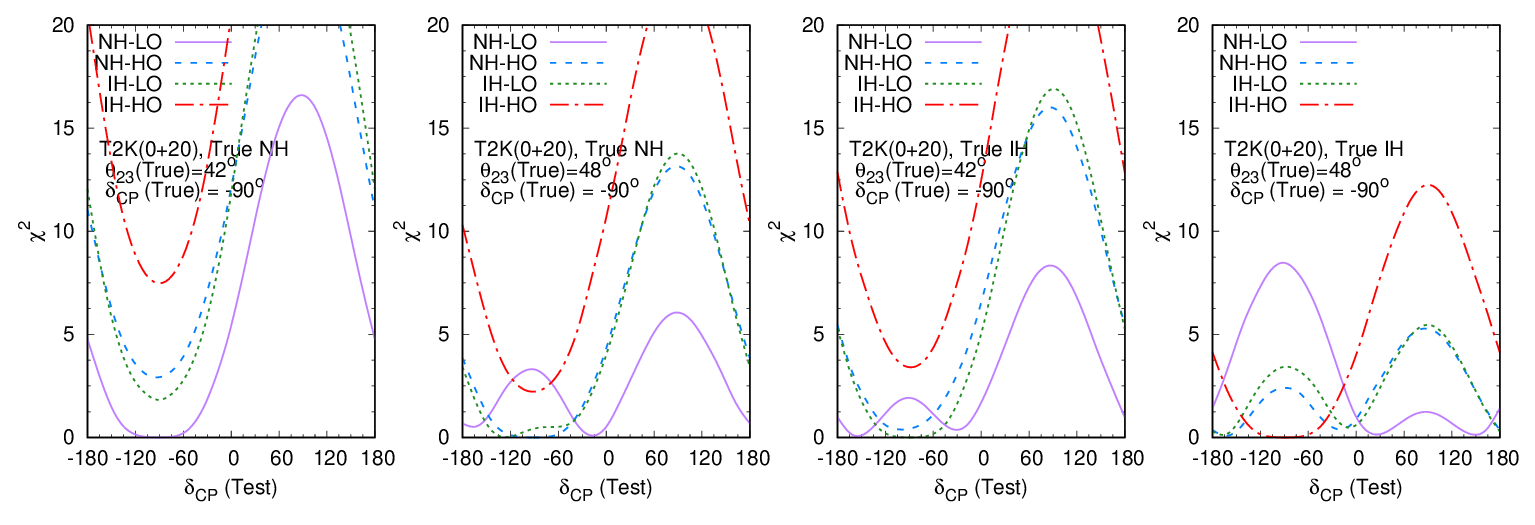}
  \includegraphics[width =\textwidth]{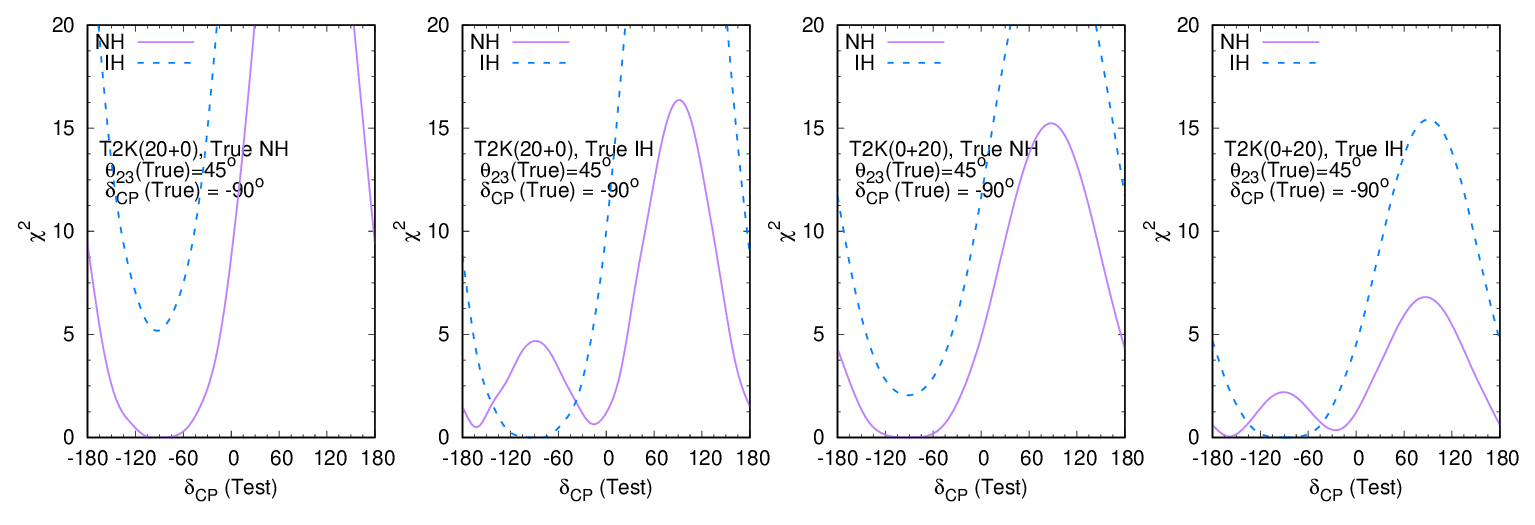}
  \includegraphics[width =\textwidth]{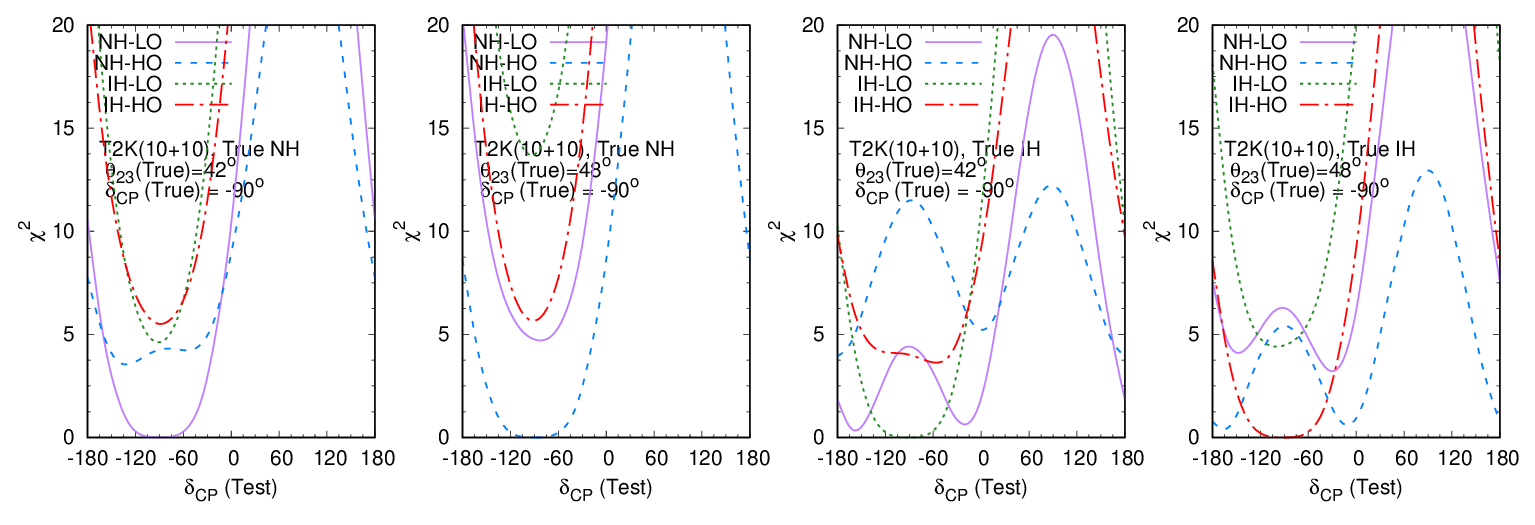}
  \caption{Hint for hierarchy and octant assuming true value of $\dcp=-90^\circ$ for an exposure of $20 \times 10^{21}$ pot. The first (second) row is for pure neutrinos (antineutrinos).
  In the third row $\theta_{23}$ is $45^\circ$ and in the fourth row, an equal neutrino+antineutrino run of T2K is assumed.
  }
  \label{cpnurun_1}
  \end{figure*}
  
  \begin{figure*}
  \epsfig{file=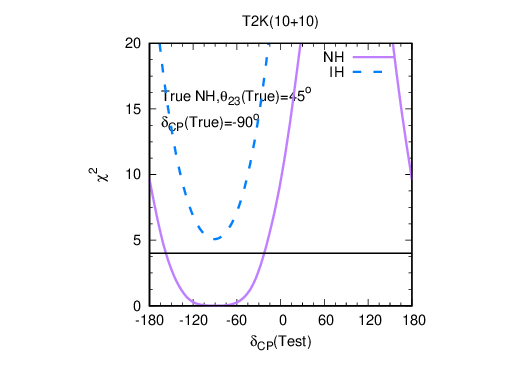, width=0.45\textwidth, bbllx=79, bblly=50, bburx=270, bbury=255,clip=} 
  \epsfig{file=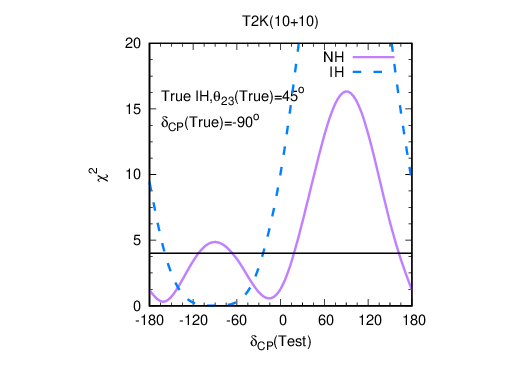, width=0.45\textwidth, bbllx=79, bblly=50, bburx=270, bbury=255,clip=} 
  \caption{$\dcp$ sensitivity of T2K for equal neutrino+antineutrino run for maximal value of $\theta_{23}$ for an exposure of $20 \times 10^{21}$ pot.}
  \label{cpantinurun_1}
  \end{figure*}

Now let us consider the implications for $\dcp=-90^\circ$
when the neutrino and antineutrino data are
added together. In the fourth row of 
Fig. \ref{cpnurun},
we have plotted the CP sensitivity taking the (4+4) configuration 
of T2K. If we compare the fourth row with the first row, then this
demonstrates the actual 
role of antineutrinos in improving the CP sensitivity of T2K. 
The first panel for true NH-LO shows that with equal neutrino-antineutrino 
data, the degenerate solutions with wrong octant -- NH-HO 
and IH-HO are excluded with $\chi^2>2$. 
Similarly for true IH-LO (third panel) the degenerate solutions 
corresponding to IH-HO get disfavoured with the antineutrino run. 
But since the antineutrinos do not help in solving hierarchy degeneracy,
the wrong hierarchy solutions corresponding to NH-LO are still allowed.  
For true IH-HO case (last panel) the degenerate solutions 
occurred for NH-HO and antineutrino run does not help in solving these. 
Rather the precision for
$\dcp=-90^\circ$, i.e. the statistical significance with which 
other $\dcp$ values can be disfavoured, 
reduces as compared 
to the full neutrino run because of less statistics.  
The plot in the second panel is for true NH-HO. 
Since there is no degeneracy, in this case also the CP sensitivity 
for $\dcp=-90^\circ$ becomes worse with antineutrino run. From these figures we also see that for both NH-LO and NH-HO, it is possible to have a clear hint
of $\dcp=-90^\circ$. Thus for a combined neutrino and antineutrino analysis, the true hierarchy turns out to be normal but the octant remains undetermined.

For the sake of completeness, in Fig. \ref{cpantinurun}, we have plotted the CP sensitivity of T2K for the (4+4) configuration but taking $\theta_{23}=45^\circ$.
This figure also shows that, similar to the pure neutrino and pure antineutrino fit, the combined neutrino-antineutrino will also imply the true hierarchy to be NH 
for maximal value of $\theta_{23}$ if nature has chosen the true value of $\dcp$ as $-90^\circ$. 

Next we discuss how far the conclusions can be 
strengthened using enhanced exposure of $20\times 10^{21}$ pot.

\subsubsection{Results for $20 \times 10^{21}$ pot}

In Figs. \ref{cpnurun_1} and \ref{cpantinurun_1}, we plot the same as that of Fig. \ref{cpnurun} and \ref{cpantinurun} but for an enhanced exposure of
$20 \times 10^{21}$ pot which is the proposed exposure
for the T2K II project. From these figures
we notice that though the nature of the curves are similar as that of earlier, 
but the significance with which the true hierarchy and true octant can be established 
from the measurement of $\dcp=-90^\circ$ is much higher. 
The 2nd panel of the first row shows that for only neutrino runs
for NH-HO 
the other wrong solutions can be disfavoured with $\Delta \chi^2 > 5$ 
(as compared to $\Delta \chi^2 \approx 3$ in the previous section). 
Whereas for only antineutrino run for true NH-LO, the other solutions 
are disfavoured with $\Delta \chi^2 \approx 2.5$ which is an improvement 
over the previous value of $\Delta \chi^2 \approx 1$ (first panel 
of 2nd row).  In all other cases, apart from these two, there 
are spurious minimas for other values of $\dcp$ and thus a signal
at $-90^\circ$ will not be possible.   
Thus with the enhanced statistics there is an increased 
contradiction between the nature of the octant when neutrino and antineutrino
data are fitted separately. 
%In this case, if neutrino and antineutrino data is fitted separately then a 
%unambiguous hint of $\dcp=-90^\circ$ will establish the hierarchy to be normal
%however there is a contradiction in the octant. 
Now we ask the question what happens if $\theta_{23} = 45^\circ$ ? 
This is addressed in the third row from where we see that for both neutrinos
and antineutrinos a clear hint at $\dcp=-90^\circ$ can come for NH. 
However the significance with which this can be established is clearly
more for neutrino run owing to higher statistics. 
IH can be disfavoured with a significance of 
$\chi^2 \approx 5 (2.5)$ for pure neutrino (antineutrino) run. This can be seen from the first and third panel of third row in Fig. \ref{cpnurun_1}. 
The corresponding numbers for the earlier
case is 2.5 and 1 respectively. 
On the other hand if the neutrino and antineutrino data are fitted together, 
then although NH is preferred for both the octants,  the wrong solutions 
are excluded with a higher significance. 
Thus for equal neutrino-antineutrino run  just from hint of $\dcp=-90^\circ$ 
the nature of octant cannot be established even with increased statistics.  
%then the neutrino data try to drive the octant towards higher and 
%and antineutrino data tries to drive it towards lower. As a result 
%both the octant remain allowed. 
  
%$\dcp=-90^\circ$ within $2 \sigma$ C.L for NH (first and second
%panels of fourth row in Fig. \ref{cpnurun_1}). For an exposure of $8 \times 10^{21}$ pot, the significance for the same is only 90\%.
Finally from Fig. \ref{cpantinurun_1} we can understand that
if true value of $\theta_{23}$ is $45^\circ$ and true value of $\dcp=-90^\circ$,then  for true NH a combined neutrino and antineutrino data of T2K will 
exclude IH 
at more than $2 \sigma$ C.L. (left panel).
Whereas with the present exposure, IH can be ruled out only at 90\% C.L.

% \underline{\it{\bf (b) Impact of antineutrino run}}: 
% Fig.~\ref{cpantinurun} is obtained using $4\times10^{21}$ pot exposure 
% each in the neutrino and antineutrino modes. This figure demonstrates the actual 
% role of antineutrinos in improving CP sensitivity of T2K. 
% As before, the true value of $\dcp$ is assumed to be $-90^\circ$. 
% The first panel for true NH-LO shows that with equal neutrino-antineutrino 
% data the degenerate solutions with wrong octant -- NH-HO 
% and IH-HO are excluded at $> 2\sigma$ confidence level. 
% Similarly for true IH-LO (third panel) the degenerate solutions 
% corresponding to IH-HO get disfavoured with the antineutrino run. 
% But since the antineutrinos do not help in solving hierarchy degeneracy,
% the wrong hierarchy solutions corresponding to NH-LO are still allowed.  
% For true IH-HO case (last panel) the degenerate solutions 
% occurred for NH-HO and antineutrino run does not help in solving these. 
% Rather the precision for
% $\dcp=-90^\circ$ i.e the statistical significance with which 
% other $\dcp$ values can be disfavoured, 
% reduces as compared 
% to the full neutrino run because of less statistics.  
% The plot in the second panel is for true NH-HO. 
% Since there is no degeneracy, in this case also the CP sensitivity 
% for $\dcp=-90^\circ$ becomes worse with antineutrino run. 

 \subsection{Discovery of CP violation} 
 
  \begin{figure*}
  \includegraphics[width =\textwidth]{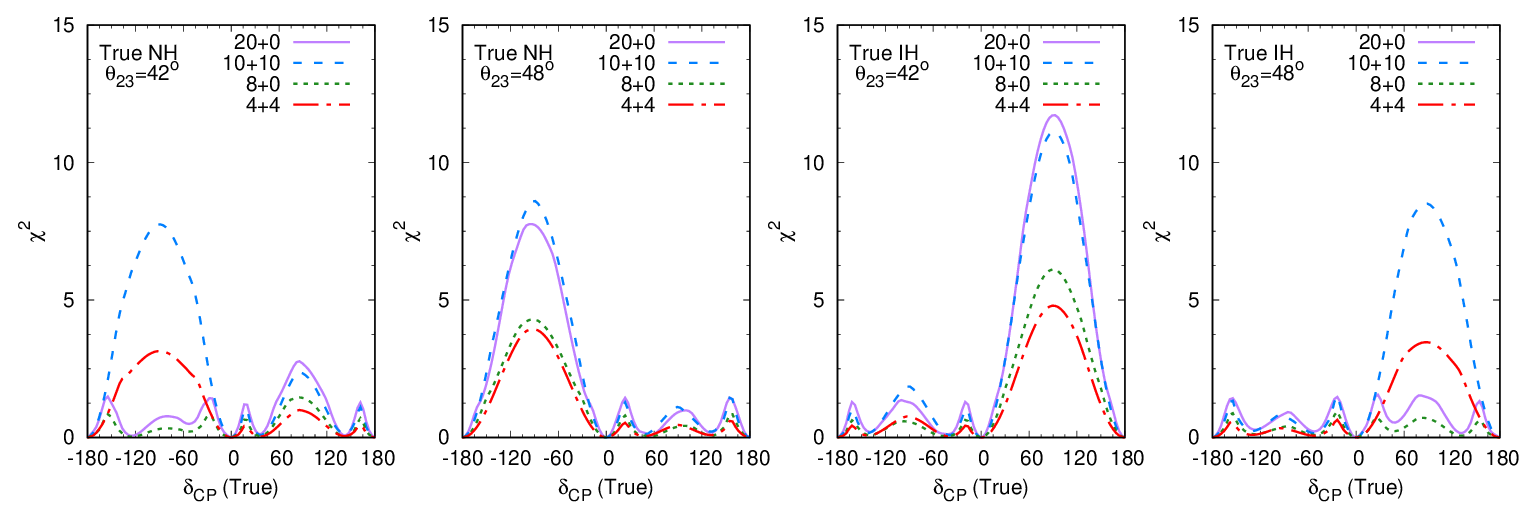}
   \caption{{$\dcp$ discovery potential of T2K for various combinations 
   of neutrino+antineutrino runs (in units of $\times 10^{21}$ pot.)
   }}
   \label{discovery}
   \end{figure*}
   
   \begin{figure*}
  \epsfig{file=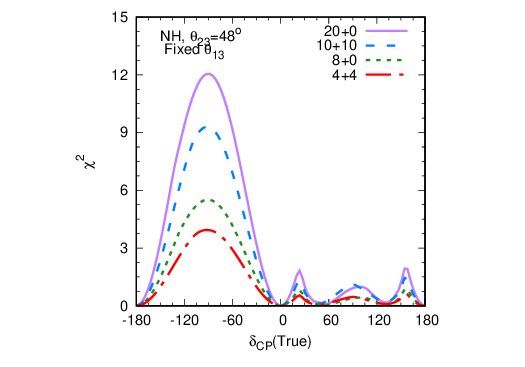, width=0.45\textwidth, bbllx=79, bblly=50, bburx=270, bbury=255,clip=} 
  \epsfig{file=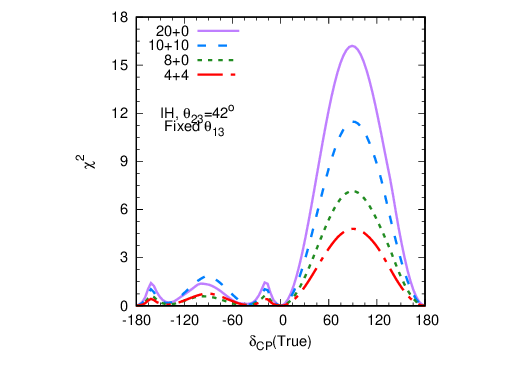, width=0.45\textwidth, bbllx=79, bblly=50, bburx=270, bbury=255,clip=} 
  \caption{$\dcp$ discovery potential of T2K for various combinations 
   of neutrino+antineutrino runs for NH-HO and IH-LO (in units of $\times 10^{21}$ pot). $\theta_{13}$ is not being marginalized.}
  \label{discovery_1}
  \end{figure*}
 
In this section we discuss the role of antineutrinos in discovering 
CP violation which implies differentiating between a true value of 
$\dcp$ from the CP conserving values $0^\circ$ or $180^\circ$. 
We present the results for different combinations
of neutrino+antineutrino exposures
-- 8+0, 4+4, 20+0 and 10+10  in units of  $10^{21}$ pot\footnote{
Note that the study of CPV discovery potential of T2K II project has been recently carried out in Ref.~\cite{Abe:2016tii} which 
compared the sensitivity between the current total exposure ($7.8 \times 10^{21}$ pot) and the proposed extended exposure ($20 \times 10^{21}$ pot) with an improved statistics 
and systematics for different
values of $\theta_{23}$. In 
our study, in the extended run we have used the same configuration as that of T2K and focused on the role of antineutrinos.
However since we have not incorporated the 
new systematics our results are somewhat worse than that in Ref.~\cite{Abe:2016tii}.
We have checked for a representative case that we get a
a 20\% improvement in our results if we incorporate the new specifications.}.
For this purpose the simulated event spectrum
is generated for true values of $\dcp$ spanning the 
range 
$[-180^\circ,180^\circ)$. This is compared with the test 
event spectrum,
with $\dcp = 0^\circ$ or $180^\circ$.  
In Fig.~\ref{discovery}, we plot the CPV discovery potential of  
T2K for different combinations of true hierarchy and octant -- NH-LO, NH-HO, 
IH-LO and IH-HO. In the test events, the hierarchy and octant 
are marginalized over. 
From the first panel, we observe that for true NH-LO no hint 
is possible at $\dcp =-90^\circ$ with only neutrino mode. 
This is because the $\chi^2_{min}$ occurs 
for test $\dcp=0^\circ, 180^\circ$ for the test NH-HO case as we 
have already seen in the top left panel of Fig. ~\ref{cpnurun}. 
 Thus ${\dcp} =-90^\circ$ cannot be distinguished from 
 ${\dcp} = 0^\circ, 180^\circ$ since the minima occurs at these values 
 for the right hierarchy-wrong octant solutions. 
% In the 
% previous section, we have seen that
Adding antineutrinos to the neutrino data resolves the wrong octant
solutions and a better sensitivity is obtained~\cite{Ghosh:2015tan}. 
In the case of IH-LO and IH-HO there are degenerate solutions 
for $\dcp = -90^\circ$ corresponding to wrong hierarchy.
Since these correspond to wrong hierarchy solutions 
adding antineutrino data is not of much help. Thus for IH the CP discovery 
potential for $\dcp = -90^\circ$ is very low because of degeneracies. 
Note that for true IH the CP  discovery  potential is high for
$\dcp= +90^\circ$.  
For true NH-HO, there are no degeneracies for 
true $\dcp=-90^\circ$ and   the 8+0 case gives a slightly  better CP
discovery potential 
than 4+4. However for the higher exposure 10+10 gives a higher 
sensitivity as compared to 20+0 for $\dcp = -90^\circ$.  We 
find that this happens because of $\theta_{13}$ marginalization.
If we repeat the same exercise for 
fixed $\theta_{13}$ then the only neutrino runs give a much better 
result as compared to equal neutrino-antineutrino run.    
This can be seen from figure \ref{discovery_1}. 

\section{Conclusions} 

In this paper we find the implications for neutrino mass hierarchy 
and octant assuming the true value of $\dcp$ chosen by nature is 
$-90^\circ$.  This is motivated by T2K data reporting a hint 
for $\dcp \sim -90^\circ$ by running {\it only in the neutrino mode} 
following the early  \nova~ neutrino run  giving similar indications. 
For this study we have focused on the T2K experiment. 
Our argument is based on the occurrence of degenerate 
solutions in the $P_{\mu e}$ probability. 
For $\dcp = -90^\circ$  there is degeneracy in 
in both $P_{\mu e}$ and $P_{\overline{\mu}\overline{e}}$ 
for IH.  On the other hand for NH 
there is degeneracy in the 
neutrino (antineutrino) probability for LO(HO).  
Thus a clear signal at $\dcp = -90^\circ$ will imply the 
hierarchy to be NH.  Now if both neutrino and antineutrino 
data point  separately towards $\dcp = -90^\circ$ then that gives rise 
to a conflicting situation for the octant since  neutrino
data should prefer LO whereas antineutrinos data would prefer HO. 
On the other hand for   $\theta_{23}= 45^\circ$  there is no such 
conflict and both neutrino data and antineutrino probabilities
can give 
a clear solution if $\dcp = -90^\circ$.   
We do analysis of T2K data assuming two exposures 
(i) $8\times10^{21}$ pot (ii) $20 \times 10^{21}$  
pot and generate simulated data of the 
experiment assuming the true $\dcp=-90^\circ$. We fit this data 
to $\dcp$ and find that a clear minima at the true value comes 
only for normal  hierarchy and $\theta_{23}$
in the higher octant if T2K runs in pure neutrino mode. 
For the other hierarchy-octant combinations, because of parameter degeneracies 
there are multiple minima which 
forbid a clear hint for $\dcp=-90^\circ$.   
Next we assume T2K running in antineutrino mode  
and repeat the above exercise. 
In this case we find that  
an unambiguous signal for $\dcp = -90^\circ$ is possible only for 
NH and LO. For the other hierarchy octant combination again there are 
more than one minima due to degeneracies. 
Thus if both neutrino and antineutrino data of T2K gives an indication 
of $\dcp=-90^\circ$ then in one case the preferred octant seems to 
be HO and in the other case it is LO. 
Thus a hint of $\dcp=-90^\circ$ in both channels would imply $\theta_{23}$ 
to be maximal.  It is interesting to note in this context that 
the best-fit $\theta_{23}$ is coming out as maximal in the T2K data~\cite{Abe:2015awa}.  
The higher exposure of T2K supports the above conclusions with a 
greater significance. 
For this work we have done the analysis with only simulated T2K data.
However we checked that similar conclusions hold true also for \nova.
%In this context it is also interesting to note that the initial \nova~
%results indeed disfavour inverted hierarchy~\cite{Adamson:2016tbq}. 

We have also examined the role of antineutrino data in the 
$\dcp$ sensitivity in T2K and  find that one of the  main contribution of 
this is to  remove the octant degeneracy. 
If this degeneracy is not present
then the decrease in statistics associated with antineutrino 
run worsens the CP sensitivity  for fixed $\theta_{13}$ 
and only neutrino run gives the better result. However if we 
marginalize over $\theta_{13}$ then the interplay between 
the neutrino, antineutrino and prior contributions 
reduces the difference in the results between pure neutrino   
runs and equal neutrino-antineutrino run and we find that for NH-HO, 10+10 is better than 20+0. 
% neutrino and  antineutrino run do not 
% always give the best results. Once the degenerate solution is eliminated
% further addition of antineutrino data worsens the CP sensitivity
% because of reduced statistics.  
The results underscore the importance of optimizing 
antineutrino run, and  
can 
significantly impact the planning of neutrino facilities in future.  
%In conclusion, we have shown for the first time  
%that if T2K continues to find a 
%hint for maximal CP violation  with $\dcp = -90^o$, 
%running in the neutrino mode then it is possible to have 
%an indication of hierarchy and octant also  
%from T2K neutrino data only. The antineutrino run is useful only when 
%octant degeneracy is present  

%If the current indication for $\dcp$ is true, then the T2K experiment 
%has the unique opportunity to give hints on all the unknown oscillation 
%parameters by running only in the neutrino mode. 
%The results also underscore the importance of optimizing 
%antineutrino run, and  
%can 
%significantly impact the planning of neutrino facilities in future.  
%Finally the knowledge about hierarchy, octant and leptonic CP phase can have 
%important ramifications 
%for neutrino model building, and hence 
%on physics beyond the Standard Model. 

%\newline

\section*{Acknowledgements}
We thank S. Uma Sankar for useful comments and discussions.   The  work
of MG is partly supported by the ``Grant-in-Aid for Scientific
Research of the Ministry of Education, Science and Culture,
Japan", under Grant No.  25105009.

%\begin{thebibliography}{000} %for 3 digits
%\begin{thebibliography}{00}  %for 2 digits

\end{document}